\documentclass[12pt]{article} 
 
\def\be{\begin{equation}}
\def\ee{\end{equation}}
\def\a{\alpha}
\def\c{\tilde{c}}
\def\b{\tilde{b}}
\def\s{\sigma}
\def\n{\{n\}}
 
\def\F{\hat{F}}
\def\vac{|0\rangle} 
\def\aa{{\large a}}
\def\ra{\rangle}
\def\la{\langle}
\def\sin{\mbox{sin}}

\begin{document}

\begin{center} 
{\large  Coordinate space wave function from the Algebraic Bethe Ansatz \\ 
for the inhomogeneous six-vertex model.} 
\end{center}

\begin{center}
{\large  A.A. Ovchinnikov  }
\end{center}

\begin{center}
{\it Institute for Nuclear Research, RAS, Moscow, 117312, Russia} 
\end{center}

\vspace{0.1in}

\begin{abstract}
We derive the coordinate space wave function for the inhomogeneous 
six-vertex model from the Algebraic Bethe Ansatz.  
The result is in agreement with the result first obtained long time ago by 
Yang and Gaudin in the context of the problem of one-dimensional fermions 
with $\delta$- function interaction.  

\end{abstract}

\vspace{0.2in}

              {\bf 1. Introduction.}

\vspace{0.2in}

The solution of the six-vertex model with an arbitrary inhomogeneity parameters 
was an important step in the theory of integrable systems.  
In the rational case the solution was given independently by Yang \cite{Yang67} 
and Gaudin \cite{G67} using the generalization of the coordinate Bethe ansatz 
(for a review see for example \cite{G}). 
Later the solution was simplified drastically in the framework of the Algebraic 
Bethe Ansatz method (for example see \cite{FST}). 
 However, since up to now the form of the coordinate space wave function of the 
eigenstates of the transfer matrix in the framework of the Algebraic Bethe Ansatz 
was not investigated, the connection of this two approaches remains obscure. 
 The goal of the present letter is to fill this gap and derive the coordinate space 
wave function of the Bethe eigenstate given by the Algebraic Bethe Ansatz. 
We find the complete agreement with the results of ref's \cite{Yang67}, \cite{G67}. 

The solution of this problem is achieved with the help of the so called Factorizing 
operator. This operator introduced in ref.\cite{MS} plays an important role in the 
solution of Quantum Inverse Scattering problem \cite{KMT}, calculations of the 
correlation functions in spin chains \cite{KMT1} and scalar products \cite{O}. 
    The calculation of the wave function which looks hopeless in terms of the 
usual Algebraic Bethe Ansatz operators, is quite simple in the F-basis for the 
operators in the auxiliary space given by the tensor product of $M$ (the number of 
up-spins) spin $1/2$ auxiliary spaces. 
    In Section 2 we introduce the notations and briefly discuss the definition and 
the properties of the Factorizing operator. We calculate the wave function of Bethe 
eigenstate in Section 3. Finally in Section 3 we present the conclusion.

\vspace{0.2in}

{\bf 2. Algebraic Bethe Ansatz and the F-basis.} 

\vspace{0.2in}

    We consider in this letter the transfer matrices corresponding to both  
rational and trigonometric regimes of the six-vertex model. 
In the present section we diagonalize the operator $A$ and introduce 
the factorizing operator.  Let us fix the notations: the normalization 
of basic S - matrix, the definition of monodromy matrix and write down the 
Bethe Ansatz equations. For the rational case the S- matrix has the 
form $S_{12}(t_1,t_2)=t_1-t_2+\eta P_{12}$, where $P_{12}$ is the permutation 
operator. In general trigonometric case it can be written as 
\[
S_{12}(t_1,t_2) =  \left( 
\begin{array}{cccc}  
a(t) & 0 & 0 & 0 \\
0 & c(t) & b(t) & 0 \\
0 & b(t) & c(t) & 0 \\
0 & 0 & 0 & a(t) 
\end{array}  \right)_{(12)}, ~~~~ t=t_1-t_2. 
\]
One can choose the normalization $a(t)=1$ so that the functions $b(t)$ and $c(t)$ 
become 
\[
\c(t)=\frac{\phi(t)}{\phi(t+\eta)},~~~  
\b(t)=\frac{\phi(\eta)}{\phi(t+\eta)}, 
\]
where $\phi(t)=t$ for the rational case and $\phi(t)=\sin(t)$ 
for the trigonometric case. With this normalization the $S$-matrix satisfies 
the unitarity condition $S_{12}(t_1,t_2)S_{21}(t_2,t_1)=1$. 
The monodromy matrix is defined as 
\[
T_{0}(t,\{\xi\})= S_{10}(\xi_1,t)S_{20}(\xi_2,t)...S_{L0}(\xi_N,t), 
\]
where $\xi_i$ are the inhomogeneity parameters and $L$ is the length of the lattice. 
We define the operator entries in the auxiliary space $(0)$ as follows: 
\[
\la \beta |T_0|\a \ra = \left( 
\begin{array}{cc}
A(t) & B(t) \\
C(t) & D(t) 
\end{array} \right)_{\a\beta}; ~~~~\a,\beta = (1,2) = (\uparrow; \downarrow). 
\]
We denote throughout the paper $(\uparrow;\downarrow)=(1;0)$ so that the 
pseudovacuum (quantum reference state) $\vac=|\{00...0\}_L\ra$. 
The triangle relation (Yang-Baxter equation) reads:  
\[
S_{12}S_{13}S_{23}=S_{23}S_{13}S_{12},~~~R_{00^{\prime}}T_0 T_{0^{\prime}}= 
T_{0^{\prime}}T_0 R_{00^{\prime}}; ~~~
R_{00^{\prime}}=S_{0^{\prime}0}. 
\]
The action of the operators on the pseudovacuum is: $A(t)\vac=\aa(t)\vac$ 
($\aa(t)=\prod_{\a}\c(\xi_{\a}-t)$), $D(t)\vac=\vac$, $C(t)\vac=0$. 
The Bethe Ansatz equations for the eigenstate of the Hamiltonian
$\prod_{i=1}^M B(q_i)\vac$ and the corresponding eigenvalue of the 
transfer - matrix $Z(t)=A(t)+D(t)$ are 
\[
\aa(q_i)=\prod_{\a\neq i} \c(q_{\a}-q_i) (\c(q_i-q_{\a}))^{-1},~~~~~ 
\Lambda(t,\{q_{\a}\})=\aa(t)\prod_{\a=1}^{M}\c^{-1}(q_{\a}-t)+ 
\prod_{\a=1}^{M}\c^{-1}(t-q_{\a}),   
\]
where $q_{\a}$ are the solution of Bethe Ansatz equations.

In the present paper we will use 
the monodromy matrix in the $F$ -basis, the basis obtained
with the help of the factorizing operator $F$ introduced in ref.\cite{MS}.   
One can construct the operator $F = F_{1...L}$ which diagonalizes the 
operator $A(t)$ \cite{MS},\cite{KMT},\cite{O} ($A^F(t)=F^{-1}A(t)F$). 
The diagonal operator $A^F(t)$ has the following 
form: 
\be
A^F(t)=\prod_{i=1}^L\left(\c(\xi_i-t)(1-n_i)+n_i\right).
\label{af}
\ee
where we denote by $n_{i}$ the operator of the number of particles 
(hard-core bosons, corresponding to the up-spin) at the site $i$.  
Let us briefly mention some of the properties of the operator $F$. 
The explicit form of the operator $F$ is 
\be
F_{12\ldots N}=\F_1\F_2\ldots \F_L,~~~~\F_i = (1-\hat{n}_i)+T_i \hat{n}_i,  
\label{f}
\ee
where $\hat{n}_i$ is the operator of the number of particles (spin up) at the 
site $i$ and the operator $T_n$ is given by the equation 
\[
T_n = S_{n+1,n}S_{n+2,n}\ldots S_{Ln}. 
\]
One can obtain the following formulas for the matrix elements of 
the operator $F$ \cite{O} in the following form: 
\[
F_{\{m\}\{n\}}=\la\{m\}|B(\xi_{n_1})B(\xi_{n_2})\ldots B(\xi_{n_M})\vac, 
\] 
where the sets of coordinates $\{m\}$ and $\{n\}$ label the positions of the 
occupied sites. The similar expression can be obtained for the inverse 
operator $F^{-1}$. Apart from diagonalizing the operator $A(t)$, the operator 
$F$ is the factorizing operator \cite{MS} in the following sense. 
For any permutation of indices $\sigma\in S_L$ 
($S_L$ - is the group of permutations) we have the equation 
$F=F^{\sigma}R^{\s}$, where $F^{\sigma}_{12..L}=F_{\s1\s2..\s L}$ 
(including the permutation of the inhomogeneity parameters $\xi_i$) 
and $R^{\s}_{1...L}$ is the operator constructed from the $S$- matrices 
defined in such a way that for the permutation of the monodromy matrix 
$T_0^{\s}=T_{0,\s1\s2..\s L}$ we have $T_0^{\s}= (R^{\s})^{-1}T_{0}R^{\s}$. 
For the particular permutation $\sigma(\n)$ such that 
$\s 1=n_1,\ldots \s M=n_M$ ($n_1<n_2< \ldots <n_M$)  
the factorization condition is represented as 
$F(F^{\sigma(\n)})^{-1}=T_{n_1}..T_{n_M}$.     
To prove the factorizing property of the operator (\ref{f}) 
it is sufficient to consider only one particular permutation, for example, 
the permutation $(i,i+1)$, since all the other can be obtained as a 
superposition of these ones for different $i$. One can show, that  
$F=S_{i+1,i}F^{(i,i+1)}$, which evidently proves the factorization property.  

The matrix elements of the operators $B(t)$ and $C(t)$  
in the F - basis: $B^{F}(t)=F^{-1}B(t)F$ (and the same for $C(t)$) 
have the following form
\be
B^{F}(t)=\sum_{i}\sigma_{i}^{\dagger}~\b(\xi_{i}-t)\prod_{k\neq i} 
\left(\c(\xi_k-t)(\c(\xi_k-\xi_{i}))^{-1}(1-n_k)+n_k \right).  
\label{bf}
\ee
\be
C^{F}(t)=\sum_{i}\sigma_i^{-}~\b(\xi_i-t)\prod_{k\neq i} 
\left( \c(\xi_{k}-t)(1-n_k)+(\c(\xi_i-\xi_{k}))^{-1} n_k\right).  
\label{cf}
\ee
The operators (\ref{bf}) and (\ref{cf}) are quasilocal i.e. they describe  
flipping of the spin on a single site with the amplitude depending on the 
positions of the up-spins on the other sites of the chain. The operator 
$D^F(t)$ can be found, for example, using the quantum determinant 
relation and has a (quasi)bilocal form.

For our calculations performed in the next Section it will be useful 
to define the operators $B_{i}(t)$, $i=1\ldots L$ creating the particle 
(spin up) at the site $i$ defined as $B^F(t)=\sum_{i}B_{i}(t)$. 
According to Eq.(\ref{bf}) we have explicitly: 
\be 
B_{i}(t)=\sigma_{i}^{\dagger}~\b(\xi_{i}-t)\prod_{k\neq i} 
\left(\c(\xi_k-t)(\c(\xi_k-\xi_{i}))^{-1}(1-n_k)+n_k \right).
\label{bi}
\ee
It is important that the operators $B_{i}(t)$ have a very simple 
commutational relations with the operator $A^{F}(t^{\prime})$ (in contrast 
to the well known commutational relations of $B(t)$ and $A(t^{\prime})$). 
In fact we have: 
\be 
B_{i}(t)A^{F}(t^{\prime})=\c(\xi_i-t^{\prime})A^{F}(t^{\prime})B_i(t). 
\label{ab}
\ee 
Equations (\ref{ab}) will be used later in the next Section.

Let us mention here that the operators $B_i=B_i(0)$ have the following 
commutational relations: 
\be 
B_{i}B_{j}=\tilde{S_{ij}}B_{j}B_{i}, ~~~~~
\tilde{S_{ij}}= \frac{\c(\xi_i)\c(\xi_j-\xi_i)}{\c(\xi_j)\c(\xi_i-\xi_j)}.
\label{aa}
\ee 
From the equation (\ref{aa}) it easy to see that the matrices $\tilde{B}_i$ 
($B_i=\tilde{B}_{i}A^F$) and $A^{F}$ taken in the auxiliary space corresponding 
to the lattice of $M$ sites $0_1,\ldots 0_M$ with the corresponding spectral 
parameters give the explicit realization of the matrices entering the so called 
Matrix Product Ansatz for the XXZ spin chain \cite{AL}.

\vspace{0.2in}

{\bf 3. Coordinate space wave function.} 

\vspace{0.2in}

We have to calculate the coordinate-space wave function for the inhomogeneous 
six-vertex model which defines the eigenstate $|\phi\ra$ of the transfer-matrix 
according to the equation 
\[
|\phi\ra=\sum_{x_1,\ldots x_M}\psi(x_1,\ldots x_M)|x_1,\ldots x_M\ra, 
\]
where the sum is over the configurations of particles (up-spins) $x_i\neq x_j$ 
for $i\neq j$, and $|x_1,\ldots x_M\ra$ is the state with the occupied sites 
$1\leq x_i\leq L $. The wave function is given by the following scalar product: 
\be 
\psi(x_1,\ldots x_M)=\la x_1,\ldots x_M|B(q_1)B(q_2)\ldots B(q_M)|0\ra,  
\label{psi}
\ee 
where $q_i$ are the parameters which obey the Bethe ansatz equations. 
We consider the wave function (\ref{psi}) in the sector 
$x_1<x_2<\ldots <x_M$.

Using the definition of the operators $B(q_i)$ in terms of the $S$- matrices, 
reordering the $S$- matrices in the product $\prod_{i}B(q_i)$, we rewrite the 
equation (\ref{psi}) in terms of the new monodromy matrices 
$T_i=S_{i0_1}S_{i0_2}\ldots S_{i0_M}$, $i=1,\ldots L$, depending on the site $i$ and 
acting in the new quantum space $(0_1,\ldots 0_M)$, as follows: 
\[
\psi(x_1,\ldots x_M)= 
\la0|\la x_1\ldots x_M|T_1T_2\ldots T_L|0\ra_{L}|\{11...1\}_M\ra, 
\]
where the second average corresponds to the space $(0_1,\ldots 0_M)$. 
We evaluate the average in the quantum space $1,2,\ldots L$, use the symmetry 
$0\leftrightarrow 1, A\leftrightarrow D, B\leftrightarrow C$ and transform the 
operators acting in the space $(0_1,\ldots 0_M)$ to the operators in the F-basis. 
One should also take into account that the action of the operator $F$ to the 
vacuum is trivial: $F|0\ra=|0\ra$.

Thus we have to calculate the following average over the states on the lattice 
with $M$ sites $0_1,\ldots 0_M$: 
\be
\psi(x_1,\ldots x_M)= 
\la\{11...1\}_{M}|A^{F}(\xi_1)A^{F}(\xi_2)...B^{F}(\xi_{x_1})...
B^{F}(\xi_{x_M})...A^{F}(\xi_L)|0\ra, 
\label{av}
\ee 
where the operators $B^{F}(\xi_{x_i})$, $i=1,\ldots M$ are located at the 
positions $x_1,\ldots x_M$ and the operators $A^F$ and $B^F$ correspond to the 
new transfer matrix of the form $T_i=S_{i0_1}S_{i0_2}\ldots S_{i0_M}$. 
The factorizing operator and the operators in the F-basis are also 
defined in the space $(0_1,\ldots 0_M)$. 
In contrast with the similar expression with the operator $A$ and $B$ in the 
usual basis, one can obtain the compact expression for the average (\ref{av}) 
with the operators in the F-basis.   

Since the matrix $A^F$ is diagonal, 
the next step to obtain the wave function is to use the equation (\ref{bi}) 
to rewrite the average (\ref{av}) as a sum over the permutations. In fact one 
obtains: 
\be
\psi(x_1,\ldots x_M)=\sum_{P\in S_M} \la\{11...1\}_{M}|A^{F}(\xi_1)
A^{F}(\xi_2)...B_{P1}(\xi_{x_1})...B_{PM}(\xi_{x_M})...A^{F}(\xi_L)|0\ra.  
\label{sum}
\ee 
Now using eq.(\ref{ab}) we commute all the operators $B_{Pi}(\xi_{x_i})$ to 
the left. The action of the operators $A^F(\xi_l)$, $l\neq x_i$ to the right- hand  
state (to the vacuum) is known, so we obtain the following compact expression 
(in the sector $x_1<x_2<\ldots <x_M$):  
\[
\sum_{P\in S_M} \prod_{l_1<x_1}\frac{1}{\c(\xi_{l_1}-q_{P1})} 
\prod_{l_2<x_2,l_2\neq x_1}\frac{1}{\c(\xi_{l_2}-q_{P2})} \ldots 
\prod_{l_M<x_M, l_M\neq x_1,...x_{M-1}}\frac{1}{\c(\xi_{l_M}-q_{PM})}
\]
\be
\prod_{l=1,l\neq x_i}^{M}\prod_{j=1}^{M}\c(\xi_l-q_j)
\la\{11...1\}_{M}|B_{P1}(\xi_{x_1})...B_{PM}(\xi_{x_M})|0\ra. 
\label{lsum}
\ee 
Note that one have to change the signs of all the spectral parameters in the 
relation (\ref{ab}) and the formula for the action of the operators $A^{F}(\xi_i)$ 
to the vacuum due to the definition of the new monodromy matrices $T_i$ acting 
in the new quantum space $(0_1,\ldots 0_M)$. 
The last step is to evaluate the average at the end of the last equation. 
Acting consequently by the operators $B_{Pi}$ to the right we obtain the 
expression: 
\[
\Phi_{M}({\xi_x},{q}|P)=\la\{11...1\}_{M}|B_{P1}(\xi_{x_1})...B_{PM}(\xi_{x_M})|0\ra 
\] 
\be
=\prod_{i=1}^{M}\b(\xi_{x_i}-q_{Pi})\prod_{i>j}\c(\xi_{x_i}-q_{Pj})
\prod_{i>j}\frac{1}{\c(q_{pi}-q_{Pj})}. 
\label{object}
\ee
The sum over the permutations $P$ of this expression gives the partition function 
of the six-vertex model with domain-wall boundary conditions 
$\sum_{P}\Phi_{M}({\xi_x},{q}|P)=\Phi_{M}({\xi_x},{q})$. 
It is interesting to obtain the determinant representation for this function 
$\Phi_{M}({\xi_x},{q})$ \cite{IK},\cite{Coker} starting from the representation 
(\ref{object}). 
In fact, representing the sum over the permutations in 
$\sum_{P}\Phi_{M}({\xi_x},{q}|P)$ as 
$\sum_{i=1}^{M}\sum_{P:PM=i}\Phi_{M}({\xi_x},{q}|P)$, and considering separately 
the dependence on the variables $\xi_{x_M}$ and $q_{PM}=q_i$, we obtain the 
following recurrence relation for the function $\Phi_{M}({\xi_x},{q})$: 
\[
\Phi_{M}(\{\xi_x\},\{q\})=\sum_{i=1}^{M}\b(\xi_{x_M}-q_i)
\prod_{\a\neq i}\frac{\c(\xi_{x_M}-q_{\a})}{\c(q_i-q_{\a})}
\Phi_{M-1}(\{\xi_{x_{\a}}\}_{\a\neq M},\{q_{\beta}\}_{\beta\neq i}), 
\]
which coincides with the well known recurrence relation which determines the 
determinant expression for $\Phi_{M}({\xi_x},{q})$ (for example, see Appendix B 
of ref.\cite{O}). 

   Substituting the expression (\ref{object}) into the equation (\ref{lsum}) 
 and performing the cancellations of similar terms we easily obtain the following result: 
\be
\psi(x_1,\ldots x_M)=\sum_{P}A(P)\phi_{P1}(x_1)\phi_{P2}(x_2)\ldots\phi_{PM}(x_M), 
\label{final}
\ee
where the functions $\phi_{Pi}(x_i)$ and the amplitude $A(P)$ are equal to 
\be
\phi_{Pi}(x_i)=\prod_{l>x_i}^{L}\c(\xi_l-q_{Pi})\b(\xi_{x_i}-q_{Pi}), ~~~~~~
A(P)=\prod_{i>j}^{M}\frac{1}{\c(q_{Pi}-q_{Pj})}.  
\label{phi}
\ee
The expression (\ref{final}) for the wave function is the final result of the 
present paper. In order to compare this result with the results obtained long time 
ago by the other authors it is useful to represent the function $\phi_{Pi}(x_i)$ 
in the following form: 
\be
\phi_{Pi}(x_i)=\left(\prod_{l=1}^{L}\c(\xi_l-q_{Pi})\right)
\left(\c^{-1}(\xi_{x_i}-q_{Pi})\b(\xi_{x_i}-q_{Pi})\right)
\prod_{l<x_i}\c^{-1}(\xi_l-q_{Pi}). 
\label{nphi}
\ee
Thus, up to the normalization factor the wave function (\ref{final}) coincides 
with the wave function obtained by the other authors (for example, see \cite{G}, 
\cite{Yang67}).

Let us obtain the Bethe ansatz equations for the inhomogeneous six-vertex model 
starting from the wave function (\ref{final}).  
We have to continue the wave function out of the interval $(1,L)$ and impose the 
periodic boundary conditions of the form: 
\be
\psi(x_1,\ldots x_M)=\psi(x_1+L,x_2,\ldots x_M)=\psi(x_2,\ldots x_M,x_1+L)  
\label{pbc}
\ee
in the sector $x_1<x_2<\ldots <x_M$. Substituting the wave function (\ref{final}) 
into the equation (\ref{pbc}) we obtain: 
\[
\sum_{P}A(P)\phi_{P1}(x_1)\ldots\phi_{PM}(x_M)= 
\sum_{P}A(PC)\phi_{P1}(x_1+L)\phi_{P2}(x_2)\ldots\phi_{PM}(x_M), 
\]
where $C$ - is the cyclic permutation ($C1=2$, $C2=3$....$CM=1$). 
This equation gives the following condition for the amplitude $A(P)$ (\ref{phi}): 
\be 
A(P)/A(PC)=\prod_{l=1}^{L}\c^{-1}(\xi_{l}-q_{P1}). 
\label{ba}
\ee 
One can easily verify that the equations (\ref{ba}) are equivalent to the 
usual Bethe ansatz equations for the six-vertex model with the inhomogeneity 
parameters $\xi_l$.

In conclusion, we derived the coordinate space wave function for the 
inhomogeneous six-vertex model from the Algebraic Bethe Ansatz. 
Our result is in agreement with the result first obtained by Yang and 
Gaudin for the eigenstate of the transfer matrix of the six-vertex model 
in the rational case.


\begin{thebibliography}{20} 


\bibitem{Yang67}
C.N.Yang, Phys.Rev.Lett, 19 (1967) 1312. 

\bibitem{G67} 
M.Gaudin, Phys.Lett.A 24 (1967) 55. 

\bibitem{G} 
M.Gaudin, ``La function d'onde de Bethe''. Masson, 1983. 

\bibitem{FST}
L.D.Faddeev, E.K.Sklyanin, L.A.Takhtajan, Theor.Math.Phys.40 (1979) 688. 

\bibitem{MS}  
J.M.Maillet and J.Sanchez de Santos, Preprint (1996), q-alg/9612012. 


\bibitem{KMT}
N.Kitanine, J.M.Maillet, V.Terras, Nucl.Phys. B 554 (1999) 647. 


\bibitem{KMT1}
N.Kitanine, J.M.Maillet, V.Terras, Nucl.Phys. B 567 (2000) 554.


\bibitem{O}
A.A.Ovchinnikov, Int.J.Mod.Phys.A 16 (2001) 2175.    


\bibitem{AL}
F.C.Alcaraz, M.J.Lazo, J.Phys.A 37 (2004) 4149.  

\bibitem{IK}
A.G.Izergin, V.E.Korepin, Commun.Math.Phys. 94 (1984) 67. \\ 
V.E.Korepin, Commun.Math.Phys. 113 (1987) 177. 

\bibitem{Coker}
A.G.Izergin, D.Coker, V.E.Korepin, J.Physics A 25 (1992) 4315. 


\end{thebibliography}
\end{document}